# Bayesian estimation of transmission networks for infectious diseases

Jianing Xu[1], Huimin Hu[1], Gregory Ellison[1], Lili Yu[2], Christopher Whalen[3], Liang Liu[1,4*]

[1]Department of Statistics, University of Georgia, Athens, GA 30606

[2]Department of Biostatistics, College of Public Health, Georgia Southern University, Statesboro, GA 30303

[3]Global Health Institute, College of Public Health, University of Georgia, Athens, GA 30606

[4]Institute of Bioinformatics, University of Georgia, Athens, GA 30602

*Corresponding author:

Liang Liu (lliu@uga.edu)

Department of Statistics

Institute of Bioinformatics

University of Georgia

310 Herty Drive, Athens, GA 30602

## Abstract

Reconstructing transmission networks is essential for identifying key factors like superspreaders and high-risk locations, which are critical for developing effective pandemic prevention strategies. In this study, we developed a Bayesian framework that integrates genomic and temporal data to reconstruct transmission networks for infectious diseases. The Bayesian transmission model accounts for the latent period and differentiates between symptom onset and actual infection time, enhancing the accuracy of transmission dynamics and epidemiological models. Additionally, the model allows for the transmission of multiple pathogen lineages, reflecting the complexity of real-world transmission events more accurately than models that assume a single lineage transmission. Simulation results show that the Bayesian model reliably estimates both the model parameters and the transmission network. Moreover, hypothesis testing effectively identifies direct transmission events. This approach highlights the crucial role of genetic data in reconstructing transmission networks and understanding the origins and transmission dynamics of infectious diseases.

# 1. Introduction

Infectious diseases remain a global health challenge, even in regions where vaccination has significantly reduced the impact of some outbreaks [1]. A crucial component of disease control is the construction and analysis of transmission networks which trace how pathogens move through populations, revealing transmission pathways between individuals [2]. By analyzing transmission networks, health professionals can identify super spreader events, pinpoint potential outbreak sources [3], and gain insights into the dynamics of pathogen transmission [4] [5] [6] [7]. Such information is vital for implementing timely interventions and developing targeted public health strategies to effectively slow or stop an epidemic's progression [8] [9] [10] [11].

Transmission networks are typically reconstructed using two main methods: contact tracing [12] [13] [14] and whole-genome sequencing (WGS) [15] [16]. Contact tracing involves systematically identifying and monitoring individuals who have been in close contact with confirmed cases of an infectious disease. While this survey-based approach is useful for tracking and controlling disease spread, it can be labor-intensive and prone to errors due to the complexity of accurately tracing interpersonal interactions [17]. In contrast, WGS provides detailed insights into the genetic variations of pathogens. By analyzing these genetic differences, WGS can map the pathways of disease transmission, identifying clusters and the direction of spread [18] [19] [20]. This method is particularly effective in environments with high mutation rates, as it can distinguish closely related transmission chains by comparing genetic sequences. When combined with epidemiological data, WGS significantly enhances the precision and effectiveness of infectious disease surveillance and control efforts.

Increasing availability of genomic data has revolutionized the field of epidemiology, providing a powerful tool for inferring transmission networks of infectious diseases [21] [22]. By analyzing the genetic variations within a pathogen's genome, phylogenetic methods can trace the paths of disease spread, identifying clusters and the direction of transmission [23] [24] [25] [26] [27]. In contrast, other methods integrate spatial and temporal data to provide a more comprehensive analysis [28] [29] [30] [31] [32]. By incorporating additional data types, these approaches have greatly enhanced the accuracy of reconstructing transmission pathways, effectively capturing both the dynamic and geographical aspects of disease spread.

However, the computational methods that use genomic data to infer transmission trees often fail to distinguish between phylogenetic trees and transmission trees, erroneously treating them as identical [33] [34]. Moreover, they frequently assume that the infection process and the pathogen's genetic evolution occur simultaneously, overlooking the genetic diversity within each host, especially in pathogens with high mutation rates and long incubation periods. These methods tend to ignore the latent period, often using the time of symptom onset as a proxy for infection time [35] [36]. To address these challenges, we first account for within-host evolution, acknowledging that genetic diversity exists within each patient. Consequently, the lineages transmitted to a susceptible individual may differ from those sampled from the transmitter at the time of infection. This assumption captures the complexity of phylogenetic trees and highlights the importance of genetic diversity in understanding transmission dynamics. Second, we introduce a Bayesian approach to infer the transmission tree directly from WGS data. This method incorporates the latent period and distinguishes between symptom onset and the actual infection time, improving the accuracy of transmission dynamics and epidemiological models. Lastly, we relax the assumption that only a single pathogen lineage is transmitted between hosts.

By allowing for the transmission of multiple lineages, our approach better reflects the complexity and variability of real transmission events. This leads to more realistic models of pathogen spread, resulting in improved epidemiological insights and more effective public health interventions.

## 2. Materials and Methods

### 2.1 The Bayesian transmission model

We develop a Bayesian model to reconstruct the transmission network of infected cases by integrating genomic data with temporal information. Temporal data consist of symptom onset times (first experience of symptoms) $T^O = \{T_i^O | i = 1, \ldots, n\}$ and removal times (i.e., recover/quarantine) $T^R = \{T_i^R | i = 1, \ldots, n\}$ for $n$ infected individuals. Genomic data, denoted as $D = \{d_1, \ldots, d_n\}$, comprise aligned pathogen genomes from $n$ infected individuals, with $d_i$ representing the genome from the individual whose symptom onset time is $T_i^O$. Both onset times and genomic sequences are ordered chronologically (i.e., $T_1^O$ is the earliest onset time and $T_n^O$ is the most recent onset time; Figure 1).

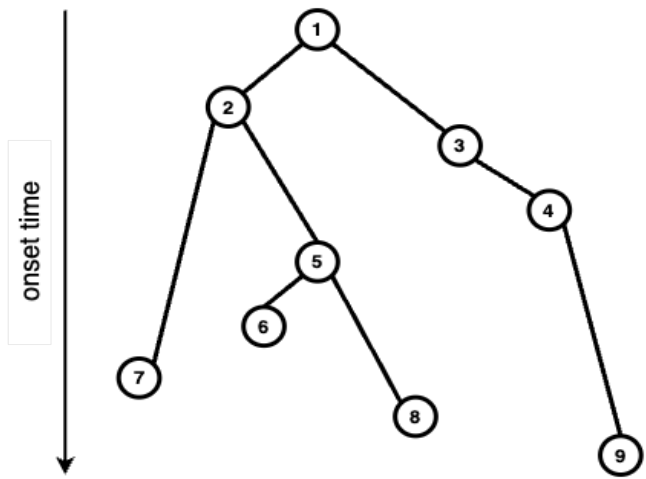

Figure 1: The transmission network for n infected individuals. The nodes represent nine infected individuals which have been ordered by their onset times. The edges represent transmission events

It is assumed that individual $i$ is infected by individual $j$ (i.e., the transmission $j \rightarrow i$) only if the onset time $T_j^O$ of individual $j$ is earlier than the onset time $T_i^O$ of individual $i$, i.e., $T_j^O < T_i^O$.

Since individuals are ordered by their onset times, $T_j^O < T_i^O$ indicates $j < i$. Thus, the transmission $j \to i$ occurs only if $j < i$. The transmission network $\Phi = \{\phi_2, \ldots, \phi_n\}$ consists of $n$ nodes (i.e., $n$ infected individuals) connected by $(n-1)$ edges $\{\phi_i | i = 2, \ldots, n\}$, which symbolize $(n-1)$ transmission events $\{J_i \to i | i = 2, \ldots, n; J_i \in \{1, \ldots, i-1\}\}$.

The Bayesian model is parametrized by the transmission network $\Phi$, the latent periods $t^L = \{t_{i,J_i}^L | i = 2, \ldots n, J_i \in \{1, \ldots, i-1\}\}$ of $(n-1)$ transmissions $\{J_i \to i | i = 2, \ldots, n; J_i \in \{1, \ldots, i-1\}\}$, the within-host effective population size parameter $\theta$, the infection rate $\alpha$, and the mutation rate $\mu$. The true infection time $T^I = \{T_i^I | i = 2, \ldots, n\}$ can be derived from the onset times and the corresponding latent periods as $T_i^I = T_i^O - t_{i,J_i}^L$. The transmission tree $\Phi$ is vital, as it documents the critical information of the host-to-host spread of pathogens. Along with within-host evolutionary parameter $(\theta)$ and the mutation rate $(\mu)$, they jointly dictate the patterns of genetic variability observed in the pathogen sequences sampled from different hosts. [37] The inclusion of the infection rate $(\alpha)$, which affects the spread dynamics, and the time information $T^I, T^R, t^L$ provide a chronological framework. Collectively, these parameters provide a comprehensive foundation for the model, accurately reflecting the interplay between evolutionary mechanisms and epidemiological patterns.

The posterior distribution $P(\Phi, t^L, \theta, \alpha, \mu | D, T^O, T^R)$ of the model parameters $\{\Phi, t^L, \theta, \alpha, \mu\}$ given the sequence data $D$ and temporal data $\{T^O, T^R\}$ is proportional to the product of the likelihood $P(D | \Phi, T^O, T^R, t^L, \theta, \alpha, \mu)$ and the prior $P(\Phi, t^L, \theta, \alpha, \mu)$, i.e.,

$$P(\Phi, t^L, \theta, \alpha, \mu | D, T^O, T^R) \propto P(D | \Phi, T^O, T^R, t^L, \theta, \alpha, \mu) P(\Phi, t^L, \theta, \alpha, \mu) \qquad (1)$$

This likelihood $P(D|\Phi, T^I, T^R, t^L, \theta, \alpha, \mu)$ represent the conditional probability of observing the sequence data $D$, given the transmission tree $\Phi$ and evolutionary parameters [38]. In addition, it is indicated that genomic data can accurately reflect transmission dynamics, ensuring that phylogenetic relationships are correctly reconstructed by integrating genetic and epidemiological information [39]. Since (n-1) transmissions $\{\phi_2, \ldots, \phi_n\}$ in the network $\Phi$ occur independently, the likelihood $P(D|\Phi, T^O, T^R, t^L, \theta, \alpha, \mu)$ is the product of the probabilities of the sequence data $D$ given (n-1) transmissions $\{\phi_2, \ldots, \phi_n\}$, i.e.,

$$P(D|\Phi, T^O, T^R, t^L, \theta, \alpha, \mu) = \prod_{i=2}^{n} P(D|\phi_i, T^O, T^R, t^L, \theta, \alpha, \mu)$$

$$= \prod_{i=2}^{n} P(d_{J_i}, d_i | \phi_i, T^O_{J_i}, T^O_i, T^R_{J_i}, T^R_i, t^L_{i,J_i}, \theta, \alpha, \mu) \quad (2)$$

In (2), $d_{J_i}$ and $d_i$ are the sequences of the individuals $J_i$ and $i$ involved in the transmission $\phi_i: J_i \to i$. Moreover, the probability $P(d_{J_i}, d_i | \phi_i, T^O_{J_i}, T^O_i, T^R_{J_i}, T^R_i, t^L_{i,J_i}, \theta, \alpha, \mu)$ can be expressed in terms of the evolutionary time $t_i$ between two sequences $d_{J_i}$ and $d_i$,

$$P(d_{J_i}, d_i | \phi_i, T^O_{J_i}, T^O_i, T^R_{J_i}, T^R_i, t^L_{i,J_i}, \theta, \alpha, \mu) = P(d_{j_1}, d_{j_2} | t_i, \theta, \alpha, \mu) \quad (3)$$

The evolutionary time $t_i$ between two sequences $d_{J_i}$ and $d_i$ is a function of $T^O_{J_i}, T^O_i, T^R_{J_i}, T^R_i, t^L_{i,J_i}$, and $\theta$, and can be expressed as below (Eqn.4) where $T^I_i = T^O_i + t^L_{i,J_i}$. Specifically, time $t_i$ equals the sum of time intervals from the time $T^{MRCA}$ of the most recent common ancestor (MRCA) of two sequences $d_{J_i}$ and $d_i$ to their removal times $T^R_{J_i}$ and $T^R_i$, i.e.,

$$t_i = (T^R_{J_i} - T^{CA}) + (T^R_i - T^{CA})$$
$$= [(T^R_{J_i} - T^I_i) + (T^I_i - T^{MRCA})] + [(T^R_i - T^I_i) + (T^I_i - T^{MRCA})]$$

$$= (t_{i,1} + t^*) + (t_{i,2} + t^*) \qquad (4)$$

$$= t_{i,1} + t_{i,2} + 2t^*$$

where $t_{i,1} = T_{j_i}^R - T_i^I$, $t_{i,2} = T_i^R - T_i^I$, and $t^* = T_i^I - T^{MRCA}$. It is assumed that the evolution of two sequences $d_{j_i}$ and $d_i$ within the host (i.e., individual $i$) is a coalescence process (Figure 2). For the comparison of two sequences $d_{j_i}$ and $d_i$, the time $t^*$ is converted to the branch length $\tau^*$ in mutation units, i.e., $\tau^* = \mu t^*$ where $\mu$ is the mutation rate. According to the coalescent theory, the branch length $\tau^*$ follows the exponential distribution with the density function $f(\tau^*|\theta) = \frac{1}{\theta} * e^{-\frac{\tau^*}{\theta}}$.

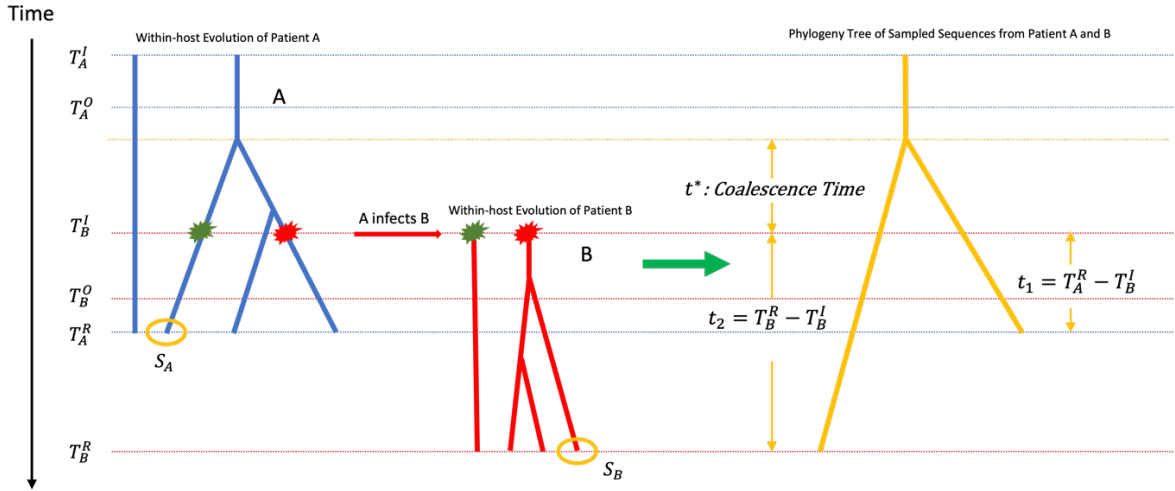

Figure 2: Illustration of within-host evolution. The blue tree represents within-host evolution of patient A, while the red tree depicts within-host evolution of patient B. The yellow tree is the coalescent tree of sampled sequences from patient A and B.

The probability $P(d_{j_1}, d_{j_2}|t_i, \mu, \theta)$ can be derived from the nucleotide substitution model. To simplify the calculation, we assume the Jukes-Cantor model for nucleotide substitutions [40]. However, any substitution model can be used to calculate the probability $P(d_{j_1}, d_{j_2}|t_i, \mu, \theta)$. Given that $\tau^*$ follows the exponential distribution with mean $\theta$, we can find the explicit expression for the probability $P(y \neq z|t_i, \mu, \theta)$ of a mutation from nucleotide $y \in \{A, C, G, T\}$ to a different type of nucleotide $z \neq y$ after time $t$,

$$P(y \neq z|t_i, \mu, \theta) = \int_0^\infty \left(\frac{3}{4} - \frac{3}{4}e^{-\mu t}\right) f(\tau^*|\theta) = \int_0^\infty \left(\frac{3}{4} - \frac{3}{4}e^{-\mu(t_{i,1}+t_{i,2})+2\tau^*}\right) * \frac{1}{\theta} * e^{-\frac{\tau^*}{\theta}} d\tau^*$$

$$= \frac{3}{4} - \frac{3}{8\theta + 4} e^{-\mu(t_{i,1}+t_{i,2})} \quad (5)$$

Let $N$ be the length (i.e., the number of sites) of the sequence alignments $D$ and $x_i$ is the number of mutations between two sequences $d_{J_i}$ and $d_i$. With the assumption of independence among sites, the probability $P(d_{J_i}, d_i|t_i, \mu, \theta)$ of two sequences $d_{J_i}$ and $d_i$ is given by

$$P(d_{J_i}, d_i|t_i, \mu, \theta) = \left(P(i \neq j|t_i, \mu, \theta)\right)^x \left(P(i = j|t_i, \mu, \theta)\right)^{N-x}$$

$$= \left(\frac{3}{4} - \frac{3}{8\theta + 4} e^{-\mu(t_{i,1}+t_{i,2})}\right)^{x_i} \left(\frac{1}{4} + \frac{3}{8\theta + 4} e^{-\mu(t_{i,1}+t_{i,2})}\right)^{N-x_i} \quad (6)$$

It follows from (6) that the likelihood function $P(D|\Phi, T^I, T^R, t^L, \theta, \alpha, \mu)$ is given by

$$P(D|\Phi, T^I, T^R, t^L, \theta, \alpha, \mu) = \prod_{i=2}^n P\left(d_{J_i}, d_i \middle| \phi_i, T^O_{J_i}, T^O_i, T^R_{J_i}, T^R_i, t^L_{i,J_i}, \theta, \alpha, \mu\right)$$

$$= \prod_{i=2}^n \left(\frac{3}{4} - \frac{3}{8\theta + 4} e^{-\mu(t_{i,1}+t_{i,2})}\right)^{x_i} \left(\frac{1}{4} + \frac{3}{8\theta + 4} e^{-\mu(t_{i,1}+t_{i,2})}\right)^{N-x_i} \quad (7)$$

Given that the population size parameter $\theta$ and the mutation rate $\mu$ have two independent priors and both parameters are independent of the transmission network $\Phi$, the latent periods $t^L$, the population size parameter $\theta$, the joint prior $P(\Phi, t^L, \theta, \alpha, \mu)$ is equal to the multiplication of three independent priors $P(\Phi, t^L, \alpha)$, $P(\theta)$ and $P(\mu)$, i.e.,

$$P(\Phi, t^L, \theta, \alpha, \mu) = P(\Phi, t^L, \alpha) P(\theta) P(\mu) \quad (8)$$

The prior $P(\theta)$ of $\theta$ is assumed to follow the exponential distribution with rate $\lambda_\theta = 1$. The prior $P(\mu)$ of the mutation rate $\mu$ is exponential with rate $\lambda_\mu = 10$. Moreover, the prior $P(\Phi, t^L, \alpha)$ is given by

$$P(\Phi, t^L, \alpha) = P(J_2 \to 2, \ldots, J_n \to n | J_2, \ldots, J_n, t^L, \alpha) P(J_2, \ldots, J_n) P(t^L) P(\alpha) \quad (9)$$

In (9), $P(\Phi | t^L, \alpha)$ is the probability of $(n-1)$ transmissions $\{\phi_i : J_i \to i | J_i \in \{1, \ldots, i-1\}; i = 2, \ldots, n\}$. Let $k_{J_i}$ be the number of individuals infected by individual $J_i$ during the infectious period $t_{J_i}^{inf} = T_{J_i}^O - T_{J_i}^R$. Equivalently, $k_{J_i}$ is the frequency of transmissions involving individual $J_i$. It is assumed that the frequency $k_{J_i}$, conditioned on $t_{J_i}^{inf}$, follows the Poisson distribution with mean $\alpha t_{J_i}^{inf}$, i.e., $P(k_{J_i} | t_{J_i}^{inf}, \alpha) = \dfrac{(\alpha t_{J_i}^{inf})^{k_{J_i}} e^{-\alpha t_{J_i}^{inf}}}{k_{J_i}!}$. The infectious period $t_{J_i}^{inf}$ follows an Exponential distribution with rate $\beta$. Given that the onset and removal times $(T^O, T^R)$ are fixed input data, $\beta$ can be reliably estimated as the mean of $T^R - T^O$ across all patients. This estimate is treated as a known constant.

Thus, the marginal distribution of $k_{J_i}$ follows a Geometric distribution with success probability $\dfrac{\beta}{\alpha+\beta}$.

$$P(k_{J_i} = n_i | \alpha) = \int_0^\infty P(k_{J_i} = n | t_{J_i}^{inf}, \alpha) f(t_{J_i}^{inf} = t) dt = \int_0^\infty \frac{(\alpha t)^n e^{-\alpha t}}{n!} \beta e^{-\beta t} dt$$

$$= \frac{\alpha^n \beta}{n!} \int_0^\infty t^n e^{-(\alpha+\beta)t} dt = \frac{\alpha^n \beta}{n!} \frac{\Gamma(n+1)}{(\alpha+\beta)^{n+1}} = \frac{\alpha^n \beta}{(\alpha+\beta)^{n+1}}$$

$$= \left(\frac{\beta}{\alpha+\beta}\right)\left(\frac{\alpha}{\alpha+\beta}\right)^n \quad (10)$$

Thus, the probability $P(J_2 \to 2, \ldots, J_n \to n | J_2, \ldots, J_n, t^L, \alpha)$ is given by:

$$P(J_2 \to 2, \dots, J_n \to n | J_2, \dots, J_n, t^L, \alpha) = \prod_{J_i} P(k_{J_i} = n_i | \alpha) = \prod_{J_i} \left(\frac{\beta}{\alpha+\beta}\right)\left(\frac{\alpha}{\alpha+\beta}\right)^{n_i} \quad (11)$$

The prior $P(t^L)$ latent periods $t^L$ is equal to the product of the probability of $(n-1)$ latent periods, i.e., $P(t^L) = \prod_{i=2}^{n} P(t_{i,J_i}^L)$ where $t_{i,J_i}^L$ follows the truncated scaled $\chi^2$ distribution with the upper bound $T_i^O - T_{J_i}^O$ and the lower bound $\max(0, T_i^O - T_{J_i}^R)$. The prior $P(J_2, \dots, J_n) = \prod_{i=2}^{n} P(J_i)$ of infectors $(J_2, \dots, J_n)$ is assumed to be discrete uniform distributions, i.e., $P(J_i) = \frac{1}{i-1}$ if no contact information is available. Given the contact probability $p_c(J_i, i)$, the probability $P(J_i)$ is proportional to the contact probability $p_c(J_i, i)$, i.e., $P(J_i) = \frac{p_c(J_i, i)}{\sum_{i=2}^{n} p_c(J_i, i)}$. In practice, the contact probability can be estimated from mobility data and geographic data. The estimated contact probability $\widehat{p_c}(J_i, i)$ can be incorporated in the Bayesian model through the prior probability $P(J_i) = \frac{\widehat{p_c}(J_i, i)}{\sum_{i=2}^{n} \widehat{p_c}(J_i, i)}$. Finally, the prior $P(\alpha)$ of the infection rate $\alpha$ is assumed to be the exponential distribution with rate $\lambda_\alpha$.

## 2.2 The Metropolis-Hastings Algorithm

Due to intractability of the posterior probability distribution $P(\Phi, t^L, \theta, \alpha, \mu | D, T^O, T^R)$, we approximate the posterior probability distribution by a sample of model parameters generated from a Metropolis-Hastings algorithm which iteratively updates model parameters until it converges to the posterior probability distribution $P(\Phi, t^L, \theta, \alpha, \mu | D, T^O, T^R)$.

The initial assignment of $J_i \in \{1, \dots, i-1\}$ (the individual who infected individual $i$) is chosen such that the corresponding sequences $d_{J_i}$ and $d_i$ contain the minimum SNPs. The initial values of the mutation rate $(\mu)$, the effective population size parameter $(\theta)$, and the infection rate $(\alpha)$ are randomly generated from uniform distributions. A random walk proposal is utilized to generate new values for the parameters $\alpha$, $\theta$, and $\mu$. To update transmission events $\{\phi_2, \ldots, \phi_n\}$ in the transmission network $\Phi$, the algorithm randomly selects a transmission event $i$ and proposes a new value for $J_i \in \{1, \ldots, i-1\}$. In addition, the latent period for the transmission $J_i \to i$ is updated by a random walk proposal similar to that used for updating the parameters $\alpha$, $\theta$, and $\mu$. In each step of the algorithm, acceptance or rejection of the proposed value is determined by the Hastings Ratio, denoted as $H$ (Supplementary Materials S1). Specifically, a random number $k$ is sampled from the uniform $[0,1]$. The acceptance probability $A$ is then calculated using the formula $A = \min(1, \exp(H))$. If $A > k$, the proposed value is accepted. Otherwise, the algorithm rejects the proposed value. The Metropolis-Hasting algorithm is implemented using Julia, which is renowned for its high-level syntax and performance akin to that of lower-level languages.

## 2.3 Hypothesis testing on direct transmissions

Comprehensively capturing all individuals within the entire transmission network for infectious disease poses significant challenges. Factors such as limited access to healthcare, the presence of undetected asymptomatic carriers, and logistical constraints in testing and data collection further contribute to these difficulties. As a result, the transmissions estimated from limited genomic data may not be direct transmissions. We have developed a statistical tool for the identification of direct transmissions. Let $d_{i\hat{J}_i}$ be the number of SNPs between two pathogen genomes. The probability distribution of $d_{i\hat{J}_i}$ is Binomial $(N, p_{i,\hat{J}_i})$, where $N$ is the total number of nucleotides in the genome and $p_{i,\hat{J}_i}$ is the probability of a mutation occurring at a locus between patient $i$ and $\hat{J}_i$. The probability $p_{i,\hat{J}_i}$ can be estimated by equation (5), i.e., $\hat{p}_{i,\hat{J}_i} = \frac{3}{4} - \frac{3}{8\hat{\theta}+4} e^{-\hat{\mu}(t_1+t_2)}$, where

$\hat{\theta}$ and $\hat{\mu}$ re the Bayesian estimates of $\theta$ and $\mu$. The mean and standard deviation of $d_{i\hat{j}_i}$ are $N\hat{p}_{i,\hat{j}_i}$ and $\sqrt{N\hat{p}_{i,\hat{j}_i}(1-\hat{p}_{i,\hat{j}_i})}$, respectively. A 95% confidence interval can be used to test if the observed number of SNPs is significantly greater than the expected number of SNPs and subsequently identify direct transmissions.

## 2.4 Simulation

The simulation of the transmission network begins with an infected individual ($i = 1$) whose infection time ($T_1^I$) is set 0. The onset time ($T_1^O$) is simulated from the scaled $\chi^2$ distribution. The removal time $T_1^R$ is simulated from the exponential distribution with rate $\beta$. Then, the number $x$ of infections caused by individual $i = 1$ during the infectious time interval $[T_1^O, T_1^R]$ is simulated from the Poisson distribution with mean $\alpha(T_1^R - T_1^O)$. The simulation terminates if $x = 0$. Otherwise, it produces new infected individuals. The infection time $T_i^I$ of the new infected individual $i$ is generated from the uniform $[T_1^O, T_1^R]$. The corresponding onset time ($T_i^O$) and remove time ($T_i^R$) are generated similarly as described above. This process continues to generate new infected individuals and corresponding infection times, onset times, and removal times. The simulation terminates when all infections exceed a predefined temporal threshold.

For each simulated transmission network, the number of SNPs between two genomes is generated from Binomial $(N, P)$ where $N$ is the genome length and $P$ is the probability of a SNP between two genomes, which is given by $P = \frac{3}{4} - \frac{3}{8\theta+4} e^{-\mu(t_1+t_2)}$ (see Equation 5). To estimate the $P_{i,J_i}$, consider three individuals $A$, $B$ and $C$ as an example, where $B$ and $C$ are infected by $A$. $P_{B,A}$ and $P_{C,A}$ can be computed by the equation below (Eqn.16). The chance that $B$ and $C$ share an identical base pair at the same locus on their TB WGS arises from two possibilities: No mutation occurred at this site among A, B and C's WGS; or mutation occurred between both $A$ and $B$, and $A$ and $C$ and the nucleotide expression at A, C are the same. This is mathematically expressed as

$P_{B,C} = 1 - [(1 - P_{B,A})(1 - P_{C,A}) + P_{B,A}P_{C,A}/3]$. All $P_{i,j}$ can be measured, allowing the estimation of the SNPs between any two given individuals.

We simulated two transmission networks with infection and removal rates of $(\alpha = 3.0, \beta = 3.0)$ and $(\alpha = 5.0, \beta = 5.0)$. For each network, we generated pairwise SNP distance matrices using four combinations of the effective population size parameter $(\theta = 10^{-6}, 5 \times 10^{-6})$ and mutation rate $(\mu = 5 \times 10^{-7}, 10^{-6})$, with a genome length of 1,000,000. We subsequently sampled 50, 100, and 200 infected individuals to obtain their pairwise SNP distances. The simulation was repeated five times.

The simulated datasets were analyzed using the Bayesian transmission model. The MCMC algorithm was run for 50,000 iterations, with the first 20,000 iterations serving as a burn-in phase. After the burn-in, parameter estimates were recorded every 100 iterations. Convergence of the algorithm was assessed through trace plots of the logarithm of the posterior probability. We also calculated the mean squared error (MSE) for each parameter ($\alpha$, $\theta$, $\mu$) across different parameter combinations and sample sizes. Let $\hat{\theta}$ be the Bayesian estimate of parameter $\theta$. The MSE of $\hat{\theta}$ is given by $MSE = \frac{1}{5}\sum_{i=1}^{5}(\hat{\theta}_i - \theta)^2$. Additionally, we assessed the accuracy of transmission network estimation by calculating the proportion of correctly inferred edges (i.e., transmission events) in the estimated network. We also measured the proportion of direct transmission events accurately identified by our proposed hypothesis test.

## 2.5 Real data analysis

Kakaire et al. [41] investigated the transmission dynamics of tuberculosis (TB) by analyzing both household and extra-household contacts of TB cases in Kampala, Uganda. The researchers enrolled 123 TB cases and 124 matched controls, with a total of 2415 first-degree network contacts identified. The samples were processed at the University of Georgia to obtain whole-genome sequencing (WGS) data. Alongside this, temporal and geographical information was meticulously documented to facilitate the reconstruction of the transmission network.

A total of 69 tuberculosis patients were selected for analysis based on the availability of complete genomic and temporal data required for the Bayesian model. The genome length for Mycobacterium tuberculosis was 4,411,532 base pairs. The MCMC algorithm was run for 5,000,000 iterations, with the first 20,000 iterations discarded as the burn-in period. After the burn-in, parameter estimates were recorded every 100 iterations. Two independent runs with different initial parameter values were conducted, and convergence was evaluated by examining the logarithm of posterior probabilities across both runs.

## 3. Results

### 3.1 Simulation

The number of infected individuals in the five networks generated for ($\alpha = 3, \beta = 3$) ranged from 242 to 769, while the infection rate ($\alpha = 5, \beta = 5$) resulted in a network with the number of infected individuals ranging from 229 to 586. All edges in the simulated networks represent direct transmission. However, sampling 50, 100, 200 infected individuals from the simulated networks resulted in smaller sub-networks in which some edges are direct while others are indirect transmissions. It appears that the proportion of direct transmissions are positively correlated the sample size (Table 1).

Table 1: The percentage of direct transmissions for different sample sizes.

| Sample Size | ($\alpha = 3, \beta = 3$) | ($\alpha = 5, \beta = 5$) |
|---|---|---|
| **50** | 17.8% | 11.3% |
| **100** | 27.9% | 25.7% |
| **200** | 56.2% | 50.8% |

To assess the performance of the Bayesian estimates for the infection rate α, the effective population size parameter θ, and the mutation rate μ, we calculated their MSEs by comparing the estimates to the true values. The MSEs for (θ, μ, α) decreased as the sample size increased from 50 to 200, and they continued to decrease with the full dataset (Figure 3A-F). At a sample size of 100, both θ and α show substantial MSE reductions of 71% and 77%, respectively, compared to a sample size of 50, demonstrating a rapid improvement in estimation accuracy (Figure 3A-F). In contrast, μ exhibits a slower reduction in MSE at smaller sample sizes. As the sample size increases to 200, all three parameters experience significant improvements in MSE, highlighting the advantage of larger datasets. With the full dataset, the MSE reductions for all parameters approach near-complete accuracy, particularly for μ, which catches up after slower initial progress (Figure 3A-F). Overall, larger sample sizes consistently lead to more accurate parameter estimates, especially for parameters that initially show slower improvement.

To determine the sample sizes needed for accurate parameter estimation, we calculated the coefficient of variation (CV) for sample sizes of 50, 100, 200, and the full dataset. At a sample size of 100, the CV for the infection rate α is 0.1, suggesting that a sample size of 100 is sufficient for accurate estimation of α. However, the full dataset is necessary to achieve a similar CV for estimating θ. In contrast, the CV for the mutation rate μ remains at 0.4 even with the full dataset, indicating less precision in estimating the mutation rate μ.

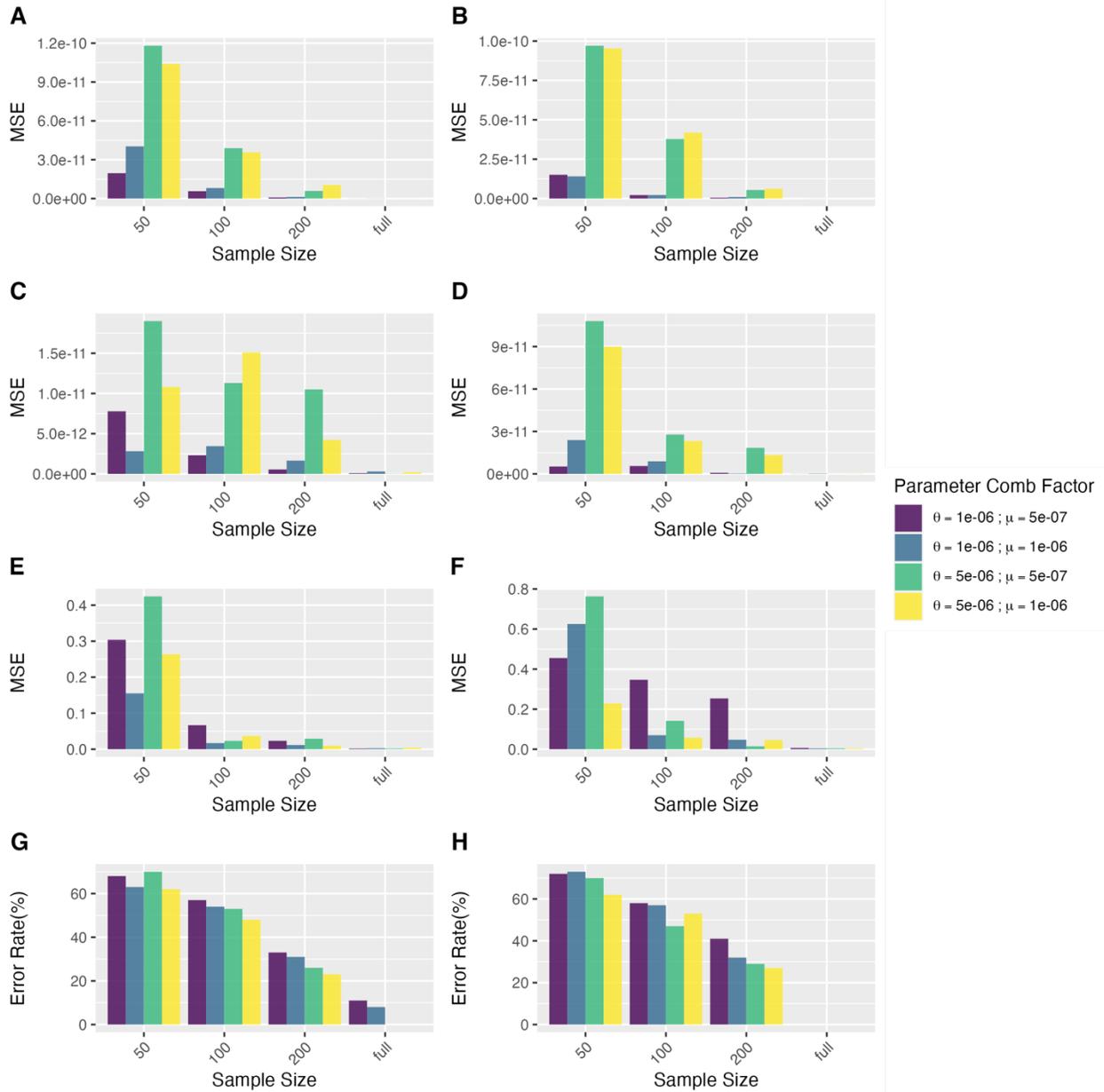

Figure 3: The MSEs of the Bayesian estimates of parameters θ, μ, α, and the transmission events in the transmission network. A) the MSE of θ for infection rate 3 ; B) the MSE of θ for infection rate 5; C) the MSE of μ for infection rate 3; D) the MSE of μ for infection rate 5; E) the MSE of α for infection rate 3 ; F) the MSE of α for infection rate 5; G) Error Rate: the proportion of wrong edges (i.e., transmission events) for infection rate 3 ; F) Error Rate: the proportion of wrong edges (i.e., transmission events) for infection rate 5;

The proportion of incorrect edges in the estimated transmission networks decreases as the sample size increases from 50 to 200, and further with the full dataset (Figure 3G-H). At smaller sample sizes, the proportion of direct transmissions in the true network is low (Table 1), suggesting that many edges in the sub-network connect patients to distant ancestors with large genetic differences, making it more challenging to accurately estimate transmission events. However, when the sample size reaches 200, the error rates for estimating transmission events drop below 30% (Figure 3G-H), indicating that a sample size of 200 is sufficient for accurately reconstructing the transmission network. Furthermore, our hypothesis testing successfully identified 100% of the direct transmissions correctly predicted by the model across various sample sizes and parameter configurations. This high level of accuracy demonstrates the model's effectiveness in accurately capturing true direct transmission events.

## 3.2 Real data analysis

In the real dataset, single nucleotide polymorphism (SNP) variation among the 69 strains ranged from 0 to 737 SNPs, with a median of 486 SNPs. The time span from the onset of symptoms in the earliest patient to the removal date of the last patient spanned 1,537 days, or approximately 4.2 years. The infectious period for these patients ranged from 14 to 731 days.

The trace plot of the log-posterior probabilities over 5,000,000 iterations after burn-in indicates that the MCMC chains have converged at the 20,000$^{th}$ iteration (burn-in) (Supplementary Materials S2). We combined the posterior samples from two independent runs. The posterior mean for the effective population size parameter $\theta$ is $1.29 \times 10^{-6}$ with the 95% credible interval of [ $5.69 \times 10^{-7}$, $2.13 \times 10^{-6}$]. For the mutation rate $\mu$, the posterior mean is $1.72 \times 10^{-6}$ with the 95% credible interval of $[1.53 \times 10^{-7}, 3.12 \times 10^{-6}]$. The posterior mean for the infection rate $\alpha$ is 4.37, with a 95% credible interval of [3.40, 5.47].

The inferred transmission network identifies patients 1 and 27 as superspreaders of the disease (Figure 4). Patient 1 transmitted the infection to 28 individuals, while patient 27 transmitted it to

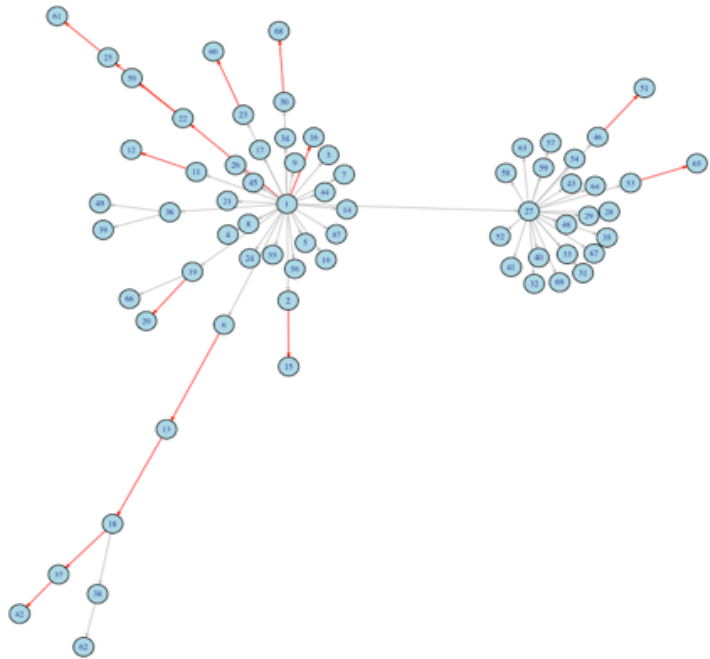

Figure 4: Transmission Network Inferred from one iteration; red edge represents inferred direct transmission; gray represents inferred indirect transmission; The number of the nodes indicates the patient ID order the ascending order the symptom onset time.

21 others (Supplementary Materials S3). Their central roles in the network emphasize their significance in the transmission dynamics and suggest the need for further investigation into their impact.

Hypothesis testing for direct transmissions was performed by comparing the observed number of SNPs to the number expected under the Bayesian model for a direct transmission. A 95% confidence interval [0, c] for the number of SNPs was constructed. If the observed SNP count for an edge in the inferred network exceeded or matched the upper bound c of the 95% confidence interval, that edge was classified as a direct transmission. The hypothesis test identified 16 direct transmissions (Table 2).

Table 2: Identification of direct transmissions in the transmission network of 69 strains. The Bayesian 95% confidence interval [0, c] for the number of SNPs associated with each of the 68 edges in the estimated network. 16 direct transmissions were identified as the observed number of SNPs was less than or equal to the upper bound c.

| Case ID | Case ID (transmitter) | Upper bound C | # of SNPs |
|---|---|---|---|
| 16 | 1 | 20.8 | 0 |
| 37 | 18 | 13.9 | 0 |
| 65 | 53 | 19.4 | 0 |
| 12 | 11 | 13.1 | 1 |
| 15 | 2 | 25.2 | 1 |
| 20 | 19 | 12.5 | 1 |
| 68 | 30 | 29.6 | 3 |
| 60 | 23 | 30.1 | 6 |
| 13 | 6 | 21.9 | 7 |
| 18 | 13 | 16.1 | 7 |
| 42 | 37 | 12.9 | 11 |
| 25 | 22 | 13.3 | 12 |
| 51 | 46 | 16.7 | 12 |
| 61 | 25 | 28.9 | 12 |
| 50 | 22 | 18.6 | 13 |
| 22 | 1 | 23.7 | 18 |

## 4. Discussion

Reconstructing transmission networks is essential for identifying critical factors like superspreaders and high-risk locations, which are key to developing effective pandemic prevention strategies. In this study, we developed a Bayesian framework that integrates genomic and temporal data to reconstruct transmission networks for infectious disease. Simulation results show that the Bayesian model accurately estimates both the model parameters and the transmission network. Additionally, hypothesis testing can successfully identify direct transmission events.

However, the limited availability of pathogen genomes can significantly reduce the performance of the Bayesian model in estimating parameters such as the infection rate, mutation rate, and effective population size. In smaller samples, the transmission network often includes many indirect transmissions. Since the Bayesian model assumes that all edges represent direct transmissions, it tends to overestimate the infection rate when indirect transmissions are present. These findings suggest that while an insufficient sample size hampers accurate estimation of the infection rate, it has minimal impact on the accurate estimation of the mutation rate and the effective population size. It's important to emphasize that the primary goal of this study is to accurately infer the transmission network. Our simulations show that the transmission network can still be reconstructed with high accuracy, even if the infection rate is not estimated precisely. Therefore, while a small sample of pathogen genomes may negatively impact infection rate estimation, it does not affect the accuracy of the transmission network reconstruction or the identification of direct transmission events.

The Bayesian model assumes a homogeneous effective population size across all hosts. A more realistic approach would relax this assumption by allowing for variable effective population sizes between different hosts. However, this would significantly increase the number of parameters, requiring a much larger sample size to accurately estimate them. Under the Jukes-Cantor model, the simplest substitution model, pairwise SNP distances are sufficient to fit the Bayesian model.

However, using a more complex model, such as the general time reversible (GTR) substitution model, the Bayesian approach can capture additional information from pathogen genomes through a sequence-based likelihood function, rather than relying solely on a genetic distance-based likelihood function. Additionally, the prior distribution for transmitters in the model is assumed to be uniform, meaning all eligible individuals are equally likely to be the transmitter. A more realistic prior would incorporate contact probabilities among individuals, which could be estimated from other types of datasets.

Bayesian analysis of transmission networks can be computationally intensive, especially for large datasets, as it involves calculating posterior distributions by integrating over the parameter space. Parallel computing provides a way to significantly reduce this computational burden by distributing the workload across multiple processors or cores. By accounting for factors that impact computational time and leveraging parallel computing techniques with appropriate software, the time required for Bayesian analysis of transmission networks can be greatly reduced, even for large datasets.

# 5. References

1. Fonkwo, P.N., *Pricing infectious disease: The economic and health implications of infectious diseases.* EMBO reports, 2008. **9**(S1): p. S13--S17.

2. Luke, D.A. and J.K. Harris, *Network analysis in public health: history, methods, and applications.* Annual review of public health, 2007. **28**(1): p. 69--93.

3. Lloyd-Smith, J.O., et al., *Superspreading and the effect of individual variation on disease emergence.* Nature, 2005. **438**(7066): p. 355--359.

4. Haydon, D.T., et al., *The construction and analysis of epidemic trees with reference to the 2001 UK foot--and--mouth outbreak.* Proceedings of the Royal Society of London. Series B: Biological Sciences, 2003. **270**(1511): p. 121--127.

5. Heijne, J.C.M., et al., *Quantifying transmission of norovirus during an outbreak.* Epidemiology, 2012. **23**(2): p. 277--284.

6. Spada, E., et al., *Use of the minimum spanning tree model for molecular epidemiological investigation of a nosocomial outbreak of hepatitis C virus infection.* Journal of clinical microbiology, 2004. **42**(9): p. 4230--4236.

7. Ypma, R.J.F., et al., *Genetic data provide evidence for wind-mediated transmission of highly pathogenic avian influenza.* The Journal of infectious diseases, 2013. **207**(5): p. 730--735.

8. Ferguson, N.M., C.A. Donnelly, and R.M. Anderson, *Transmission intensity and impact of control policies on the foot and mouth epidemic in Great Britain.* Nature, 2001. **413**(6855): p. 542--548.

9. Keeling, M.J., et al., *Modelling vaccination strategies against foot-and-mouth disease.* Nature, 2003. **421**(6919): p. 136--142.

10. Wallinga, J. and P. Teunis, *Different epidemic curves for severe acute respiratory syndrome reveal similar impacts of control measures.* American Journal of epidemiology, 2004. **160**(6): p. 509--516.

11. Heijne, J.C., et al., *Enhanced hygiene measures and norovirus transmission during an outbreak.* Emerg Infect Dis, 2009. **15**(1): p. 24-30.

12. Mollentze, N., et al., *A Bayesian approach for inferring the dynamics of partially observed endemic infectious diseases from space-time-genetic data.* Proceedings of the Royal Society B: Biological Sciences, 2014. **281**(1782): p. 20133251.

13. Worby, C.J., et al., *Reconstructing transmission trees for communicable diseases using densely sampled genetic data.* The annals of applied statistics, 2016. **10**(1): p. 395.

14. Snitkin, E.S., et al., *Tracking a hospital outbreak of carbapenem-resistant Klebsiella pneumoniae with whole-genome sequencing.* Science translational medicine, 2012. **4**(148): p. 148ra116--148ra116.

15. Ypma, R.J.F., W.M. van Ballegooijen, and J. Wallinga, *Relating phylogenetic trees to transmission trees of infectious disease outbreaks.* Genetics, 2013. **195**(3): p. 1055--1062.

16. Liu, J., et al., *SARS transmission pattern in Singapore reassessed by viral sequence variation analysis.* PLoS medicine, 2005. **2**(2): p. e43.

17. Gardy, J.L., et al., *Whole-genome sequencing and social-network analysis of a tuberculosis outbreak*. New England Journal of Medicine, 2011. **364**(8): p. 730-739.

18. Ruan, Y., et al., *Comparative full-length genome sequence analysis of 14 SARS coronavirus isolates and common mutations associated with putative origins of infection*. The Lancet, 2003. **361**(9371): p. 1779--1785.

19. Klinkenberg, D., et al., *Simultaneous inference of phylogenetic and transmission trees in infectious disease outbreaks*. PLoS computational biology, 2017. **13**(5): p. e1005495.

20. Didelot, X., J. Gardy, and C. Colijn, *Bayesian inference of infectious disease transmission from whole-genome sequence data*. Molecular biology and evolution, 2014. **31**(7): p. 1869--1879.

21. Kser, C.U., et al., *Rapid whole-genome sequencing for investigation of a neonatal MRSA outbreak*. New England Journal of Medicine, 2012. **366**(24): p. 2267--2275.

22. Eyre, D.W., et al., *A pilot study of rapid benchtop sequencing of Staphylococcus aureus and Clostridium difficile for outbreak detection and surveillance*. BMJ open, 2012. **2**(3): p. e001124.

23. Fraser, C., et al., *Pandemic potential of a strain of influenza A (H1N1): early findings*. science, 2009. **324**(5934): p. 1557--1561.

24. Harris, S.R., et al., *Evolution of MRSA during hospital transmission and intercontinental spread*. Science, 2010. **327**(5964): p. 469--474.

25. Walker, T.M., et al., *Whole-genome sequencing to delineate Mycobacterium tuberculosis outbreaks: a retrospective observational study.* The Lancet infectious diseases, 2013. **13**(2): p. 137--146.

26. Pybus, O.G. and A. Rambaut, *Evolutionary analysis of the dynamics of viral infectious disease.* Nature Reviews Genetics, 2009. **10**(8): p. 540--550.

27. Didelot, X., et al., *Transforming clinical microbiology with bacterial genome sequencing.* Nature Reviews Genetics, 2012. **13**(9): p. 601--612.

28. Lau, M.S.Y., et al., *A systematic Bayesian integration of epidemiological and genetic data.* PLoS computational biology, 2015. **11**(11): p. e1004633.

29. Didelot, X., et al., *Genomic infectious disease epidemiology in partially sampled and ongoing outbreaks.* Molecular biology and evolution, 2017. **34**(4): p. 997--1007.

30. Hall, M., M. Woolhouse, and A. Rambaut, *Epidemic reconstruction in a phylogenetics framework: transmission trees as partitions of the node set.* PLoS computational biology, 2015. **11**(12): p. e1004613.

31. Campbell, F., et al., *outbreaker2: a modular platform for outbreak reconstruction.* Bmc Bioinformatics, 2018. **19**: p. 1--8.

32. Didelot, X., et al., *Bayesian inference of ancestral dates on bacterial phylogenetic trees.* Nucleic acids research, 2018. **46**(22): p. e134--e134.

33. Nbel, U., et al., *A timescale for evolution, population expansion, and spatial spread of an emerging clone of methicillin-resistant Staphylococcus aureus.* PLoS pathogens, 2010. **6**(4): p. e1000855.

34. Mutreja, A., et al., *Evidence for several waves of global transmission in the seventh cholera pandemic.* Nature, 2011. **477**(7365): p. 462--465.

35. De Maio, N., C.-H. Wu, and D.J. Wilson, *SCOTTI: efficient reconstruction of transmission within outbreaks with the structured coalescent.* PLoS computational biology, 2016. **12**(9): p. e1005130.

36. Jombart, T., et al., *Bayesian reconstruction of disease outbreaks by combining epidemiologic and genomic data.* PLoS computational biology, 2014. **10**(1): p. e1003457.

37. Samuel, A., L. Fabio, and R.R. Roland, *Epidemiological and clinical consequences of within-host evolution.* Trends in Microbiology, 2011. **19**(1): p. 24-32.

38. Rannala, B. and Z. Yang, *Probability distribution of molecular evolutionary trees: A new method of phylogenetic inference.* Journal of Molecular Evolution, 1996. **43**(3): p. 304-311.

39. Cottam, E.M., et al., *Molecular epidemiology of the foot-and-mouth disease virus outbreak in the United Kingdom in 2001.* Journal of Virology, 2006. **80**(22): p. 11274-11282.

40. Jukes TH, C.C., *Evolution of Protein Molecules.* 1969, Academic Press: New York. p. 21-132.

41. Kakaire, R., et al., *Excess Risk of Tuberculosis Infection Among Extra-household Contacts of Tuberculosis Cases in an African City.* Clin Infect Dis, 2021. **73**(9): p. e3438-e3445.